\documentclass{PoS}

\title{Teaching particle physics to high school teachers}

\ShortTitle{Teaching particle physics}

\author{\speaker{Nils-Erik Bomark}\\%\thanks{A footnote may follow.}\\
        University of Agder, Norway\\
        E-mail: \email{nilseb@uia.no}}

\abstract{In Norway, particle physics is part of the high school curriculum in physics which introduces the need for good university teaching in particle physics without the usual technical approach.\\
Given how much conflicting information and inaccurate explanations there are on the
subject; how should we teach this to people without much knowledge in mathematics? \\ By carefully explaining the fundamental consepts of the theory it is fully possible to achieve an appreciation of particle physics without much mathematics. Through the use of analogies, such as an analogy between the freedom in choosing timezone and the freedom in choosing phase angle, one can introduce gauge theory and hence show the underlying structure of the standard model.}

\FullConference{EPS-HEP 2017, European Physical Society conference on High Energy Physics\\
		5-12 July 2017\\
		Venice, Italy}

\begin{document}

\section{The challenge of teaching particle physics}

Particle physics is perhaps the most intriguing topic in science, and captures the imagination of anyone with an interest in physics. However, the need for simplification, perhaps in combination with carelessness on the part the authors, has rendered the popular description of the field largely flawed and in many cases simply wrong.

This is not restricted to just particle physics, many of the misrepresentations stem from quantum mechanics. Perhaps the most common misunderstanding regards the interpretation of the Heisenberg uncertainty relation between energy and time, $\Delta E \Delta t \geq \hbar/2$, which is often claimed to mean that ``we can borrow some energy from the vacuum if we just pay it back fast enough''.  In the above statement $\Delta E$ is given the meaning of a change in energy, this is clearly wrong since $\Delta E$ is always the standard deviation of some energy distribution (and should perhaps therefore be denoted $\sigma_E$ rather than $\Delta E$). One should point out here that energy is always strictly conserved in all of physics, including processes with virtual particles.

To make matters worse, the inaccuracies are not restricted to popular physics, they extend through high school and all the way into university level physics books. One common example of this is the presentation of Yukawa's introduction of mesons in 1935 (see e.g.~\cite{UP}). This starts with the claim that to have a force with a range of $10^{-15}$ m, we need a particle that has $c\tau\approx 10^{-15}$ m, where $\tau$ is the particle's lifetime. One so uses the above mentioned incorrect interpretation of $\Delta E \Delta t \geq \hbar/2$ to argue that these particles should have a mass around 100-200 MeV.

The problems with the above arguments are not hard to spot, most obviously, mesons have lifetimes many orders of magnitude larger than the above arguments would suggest (they show up in detectors at the LHC). It is also rather strange to put a mass into $\Delta E \Delta t \geq \hbar/2$. Of course, Yukawa himself did nothing like this but constructed a field theory that reproduces the potential observed in experiments and in the process needed to introduce a mass term of 200$m_e$, which after quantization gives particles of that mass~\cite{Yukawa:1935xg}. As a matter of fact, a simple but correct argument would be to note that we need a length scale of a bit over $10^{-15}$m and since $10^{-15}{\rm m}\approx 1/(200\ {\rm MeV})$, we need a field whose quanta will have mass of that scale.

\section{Quantum field theory, particles and virtual particles}

One of the major confusions in quantum mechanics, is the concept of particles. Quantum field theory can here come to the rescue and make life simpler for a change. By emphasising that the fundamental objects of nature are fields, and that particles are quantized oscillations in those fields, it should be clear that when we say particle, we are talking about quantization effects and not some classical solid ball like most students (and probably teachers) imagine.

Since no joy can last forever, we need to talk about virtual particles too. This is a case of bad naming, one should emphasise that they are not particles in any sense (not the above either), but (as described by~\cite{Strassler}) different types of disturbances of the fields. In a Feynman-diagrammatic sense one can see them as mathematical tools for calculating probabilities.

\section{The importance of gauge theory}
Since the details of the standard model of particle physics are rather technical, on a lower level it easily becomes mostly learning particle names and interactions. The key to a deeper understanding and appreciation of the construction, is gauge theory.

Although gauge theory is the basis of all modern physics, it is strangely absent in most of physics curriculum, usually only properly discussed in higher level quantum field theory or particle physics courses. This is strange since gauge theory is not all that difficult; compared to teaching quantum field theory, gauge theory is a cake-walk.

The first step in introducing gauge theory to students without advanced physics background, is to carefully explain what we mean by a symmetry. Show examples of how transforming the mathematical objects of the theory does not alter the physics. One example here is how changing the scalar potential $V\to V+V_0$ for any constant $V_0$ does not change $\vec E$, and how $A_\mu\to A_\mu+e\partial_\mu\phi(x_\mu)$ for any function $\phi(x_\mu)$ does not change neither $\vec E$ nor $\vec B$; this should serve as familiar examples (at least the first one) that plants the idea that forces and symmetries have some connection.

\section{A simple analogy for a U(1) gauge theory}
In a class unfamiliar with lagrangians, gauge theory still poses a challenge. A good analogy to tackle this uses timezones as an example symmetry\footnote{The origin of this analogy is not clear, it is discussed in~\cite{Gaugeland} but any more formal reference is missing. Another interesting gauge theory analogy is using an economic system~\cite{Maldacena:2014uaa}.}. 

It should be obvious that the choice of timezone should not impact any physics experiment and changing timezone therefore represents a symmetry.

The next step is to gauge the symmetry, i.e.\ insist that we should be allowed to use different timezones in different points in space. If this is to work we clearly need something that tells us how the timezone is changing from point to point. This something needs to tell us how the timezone is changing in all three directions and hence needs to be a vector-field. Thus we have concluded that a local symmetry requires a vector-field, $\vec A$, and since this field should tell us the change in timezone, it has to transform like $\vec A\to \vec A +\nabla \phi(\vec x)$ if we change all timezones with an amount $\phi(\vec x)$.

The similarity between $\vec A$ and $A_\mu$ should now be clear and we have established electromagnetism as a gauge theory. There is a weak point in why the vector-field should be allowed to be dynamic, but that is usually not touched in higher level courses either.

This analogy has been used with success in the class-room, the conclusion being that gauge theory can be introduced without much mathematics but more effort needs to be spent on introducing the concept of symmetry.

Due to the increased mathematical complexity of the $SU(2)$ and $SU(3)$ symmetries, it is not realistic to find a good analogy for them, but since the connection between symmetry and force is now made, one can simple introduce them as they are and state that they give rise to forces too when gauged.

%\subsection{The Higgs mechanism}

\section{Conclusions}

Particle physics is a mathematically complicated subject that to a large extent is reserved for the few who spend years at universities studying it.

However, by carefully introducing consepts like particles, virtual particles and feynman diagrams, it is possible to convey the essence of the field without the usual mathematical complexity. 

By using suitable analogies one can also introduce the consept of gauge theory and hence it is possible to get a good understanding of the structure of the standard model.

\end{document}